# Blockchain and Machine Learning for Fraud Detection: A Privacy-Preserving and Adaptive Incentive Based Approach

**Tahmid Hasan Pranto[1], Kazi Tamzid Akhter Md Hasib[1], Tahsinur Rahman[2], AKM Bahalul Haque[3], A.K.M. Najmul Islam[3] and Rashedur M. Rahman[1], Senior Member, IEEE**

[1]Department of Electrical and Computer Engineering, North South University, Dhaka-1229, Bangladesh
[2]Department of Computer Science and Engineering, Brac University, Dhaka-1212, Bangladesh
[3]Software Engineering, LENS, LUT University, Finland

Corresponding author: AKM Bahalul Haque (e-mail: bahalul.haque@lut.fi).

"This work was supported by the Foundation for Economic Education (www.lsr.fi)."

**ABSTRACT** Financial fraud cases are on the rise even with the current technological advancements. Due to the lack of inter-organization synergy and because of privacy concerns, authentic financial transaction data is rarely available. On the other hand, data-driven technologies like machine learning need authentic data to perform precisely in real-world systems. This study proposes a blockchain and smart contract-based approach to achieve robust Machine Learning (ML) algorithm for e-commerce fraud detection by facilitating inter-organizational collaboration. The proposed method uses blockchain to secure the privacy of the data. Smart contract deployed inside the network fully automates the system. An ML model is incrementally upgraded from collaborative data provided by the organizations connected to the blockchain. To incentivize the organizations, we have introduced an incentive mechanism that is adaptive to the difficulty level in updating a model. The organizations receive incentives based on the difficulty faced in updating the ML model. A mining criterion has been proposed to mine the block efficiently. And finally, the blockchain network is tested under different difficulty levels and under different volumes of data to test its efficiency. The model achieved 98.93% testing accuracy and 98.22% Fbeta score (recall-biased f measure) over eight incremental updates. Our experiment shows that both data volume and difficulty level of blockchain impacts the mining time. For difficulty level less than five, mining time and difficulty level has a positive correlation. For difficulty level two and three, less than a second is required to mine a block in our system. Difficulty level five poses much more difficulties to mine the blocks.

## I. INTRODUCTION

Financial fraud has been a serious matter of concern in the business community. Even with technological advancement, fraud cases are on the rise [1]. According to the recent Juniper Research report [2], merchants are expected to suffer a cumulative loss of $206 billion due to online transaction fraud within the years 2021-2025. Robust fraud detection algorithmic model development requires authentic data. However, due to the confidential nature of financial data, privacy concerns, and lack of synergy between organizations, financial transaction data is rarely available for machine learning research [3, 4]. On top of that, organizations often abstain from collaborative research on financial cases to conceal their financial statements and business strategies [4]. Large enterprises around the world own maximum of systems and services, and a massive amount of data is also at their possession [5]. Researchers from small organizations often rely on crowdsourcing to acquire data, which is susceptible to spoofing and often proves to be completely unusable [6]. On the contrary, isolated and proprietary intra-organization solutions often become biased to single evidence where an ML model is trained only with intra-organizational data and often fails to represent the overall distribution in real-world systems [7]. For Machine Learning (ML) systems to be reliable and accurate, authentic real-world data is necessary. However, such data are rarely available due to privacy concerns, as we explained earlier.

The traditional batch-trained machine learning techniques prove to be inharmonious with the online use cases such as transaction fraud detection [8]. High-traffic platform such as e-commerce deal with a high volume of data flow. As a result, an offline ML model (batch-trained) becomes incapacitated over time and requires a manual update, which is computationally costly, laborious and impractical in an online scenario [9]. On top of that, batch-learning approaches lack the capability to incorporate insights from new data instances in a deployed (online) ecosystem [10]. Incremental machine learning is a potential solution that can aid the purpose of keeping the model updated over time, minimize the computation cost, and avoid manual updates [11]. ML algorithms learn from a data stream in an incremental machine learning approach while the model is already in service [10]. Yet, to incrementally build a better fraud detection ML model, a continuous supply of data is essential so that the model is consistent with the new trends in fraudulence. This study proposes the solution to privacy issues in building an ML model incrementally using blockchain and smart contract technology.

Research work by Tiernanm Barry provides empirical evidence that, in the online ecosystem, batch-trained ML models become imprecise and unreliable over time. In contrast, incremental ML models evolve over time and adapt to new data [9]. His study on predicting the next-minute price of three major crypto-currencies reveals that incremental online learning delivers the best performance, beating batch learning. The study by Chao Yang proposes a blockchain-based incremental outlier clustering technique that is computationally less expensive and more effective in terms of performance [12]. Blockchain technology is progressively being adopted to enhance system security while machine learning techniques are used for their predictive capabilities [13].

Taking advantage of both these technologies (ML and blockchain), Justin D. Harris and Bo Waggoner proposed a platform for collaborative machine learning [14] where a combination of incentive mechanism and data handler provides the functionality to train a machine learning algorithm on the fly collaboratively. However, their approach did not incorporate the case of corporate privacy, adaptive incentive, and specific mining criteria, [14] which are the vital aspects of the collaborative and distributed ecosystem. Furthermore, privacy is an uncompromisable aspect in businesses like e-commerce, where data contains business insights and the personal information of the consumers [15]. Considering these issues, our paper proposes a novel blockchain and smart-contract-based privacy-preserving solution for e-commerce where a fraud detection model is incrementally trained, maintained and updated.

The feasibility of blockchain technology and smart contracts in collaborative and distributed machine learning has already been studied [16]. To ensure safe collaboration, privacy preservation is preliminary. Our approach comprises testing several machine learning algorithms on an initial dataset to select one algorithm. After the initial selection, the machine learning algorithm is stored within a blockchain network to be used by any organization and to be incrementally upgraded. Smart contracts are stored within the blockchain so that the contract itself is immutable and unalterable. These contracts serve the best version of the ML model, maintain the incentive mechanism and deal with the model update calculations automatically.

To ensure data privacy, sophisticated data provided by the organizations is not shared in the central storage. Instead, the model training is done off-chain (within the organization), and only the metric is stored in the system. While incrementally updating the model by the system, the smart contract will always point to the current best model which is made available to the users. A combination of precision, recall, f-score and false-negative rate is used to update the current best ML model with a newer version. A difficulty-adaptive incentive mechanism has also been engineered to encourage the organizations so that their efforts are awarded fairly. The more difficult it is to update a model, the more incentive is disbursed. As ML, blockchain and smart contract-like cutting-edge technologies are being used together, the efficiency of the overall system has been tested under varying difficulty levels and data volume. Finally, blockchains' cryptographic distributed ledger technology secures all the information stored. To the best of our knowledge, no study has yet contemplated the case of collaborative, incremental financial fraud detection while ensuring privacy and security using smart contracts and blockchain technology. Our main contributions are summarized as follows.

- We proposed a privacy-preserving approach so that business organizations (e-commerce in this case) can build robust ML models incrementally.
- Our study proposes a difficulty adaptive incentive mechanism that incentivizes the contributor based on the level of difficulty to update a machine learning model with a newer version.
- To update a model with a newer version, a mining criteria has been proposed which utilizes a combination of model metrics to take the decision of mining a block. The mining criteria have been designed explicitly for high-traffic real-world systems like e-commerce.
- We have experimented with the mining time to test the system's efficiency for different blockchain mining difficulty levels and with different volumes of data to train the models.

The rest of the paper is documented as follows. Section 2 contains the background study on our topic, delineating the research progression. Section 3 describes our machine learning model, dataset and feature selection. This section also explains the selection procedure of the ML model and the evaluation criteria. While section 4 provides a detailed architecture of our system, section 5 exhibits the implementation details. The results of system testing have been shown in section 6. After a discussion about the system implications in section 7, the paper is concluded in section 8.

## II. Background Study

### A. Blockchain & Smart Contract

Blockchain is a progressive technology that has gained attention from the research and business communities after the massive success of bitcoin. Blockchain is a distributed ledger technology that stores the records of transactions or data using cryptography [17]. The records of transactions or data are stored as blocks of information across a peer-to-peer network [18]. The very first block is known as a genesis block, and each block is connected to its previous block [19].

The essence of blockchain is to keep transaction records decentralized, immutable, transparent, available, and secure while ensuring anonymity [18]. Blockchain is a decentralized peer-to-peer network that is not bound to a central authority, and the control over the data is not concentrated towards any node or group of nodes; instead, all the connected nodes share the same amount of authority over the blockchain network [20]. A key characteristic of blockchain is immutability which indicates preventing data alteration. Blockchain maintains an increment-only digital ledger-- data cannot be edited or deleted once added to the network [21]. In a blockchain, every node connected to that network possesses a copy of the current ledger [17]. The data inside the blockchain network is accessible by all connected participants, and this data is always available, making the system transparent and available [22]. Apart from these aspects, blockchain is also acclaimed to be robustly secure.

With its unique characteristics, blockchain has been used to solve many drawbacks of traditional systems [23, 72]. The finance sector has seen the breakthrough implementations of blockchain [24]. Blockchain solves the dilemma of double-spending (trying to use a digital currency that has already been used) [25] by keeping timestamps for every transaction under cryptographic security, along with its transaction identity. On top of that, the distributed consensus mechanism makes sure that most of the nodes have approved the validity of the new transaction [18]. Iqbal and Matulevicˇius further explored these concerns and proposed a framework for the early detection of double-spending [25]. Blockchain-based solutions establish highly secure and trustworthy mediums for business and trade cases like insurance, investment deals, venture capital financing, etc. [24]. For instance, Roriz and Pereira developed a blockchain-based solution for vehicle insurance fraud like double-dipping (multiple insurances for the exact vehicle) [26]. Apart from the financial sector, this technology has been widely adopted in sectors like healthcare [27], education [28], process automation [29] and supply chain management [30].

Although blockchain provides variegated beneficial attributes, integrating blockchain with real-world applications needs additional functionalities to represent complex transactions [18]. This is where smart contracts are introduced to the blockchain network. Smart contracts are digitally written contracts that can self-execute when a specific term is satisfied [31]. By deploying the contract inside the blockchain network, the contract becomes immutable and contractual breach becomes more restrictive and inconvenient to attackers [32]. Smart contract can sometimes invoke fraudulence, and studies have also shown how to detect those. The study of Liu et al. proposes a potential solution to financial fraud on the Ethereum blockchain [33]. In their study, the feature is extracted from complex hierarchical information in smart contracts and these features are represented as a relationship matrix. A transformer network learns an embedding from this matrix which works as the input of a classification network which then predicts financial fraud.

Sharma et al. [34] investigate the feasibility of using blockchain and smart contract for medical big data. The system proposed in their work uses blockchain to maintain the security and integrity of data, and smart contract administers the user authorization and data sharing policy and works as a controller to the overall network [34]. Tan et al. [35] highlight the practicable usage of blockchain and smart contracts to cross-organizational SLA (service legal agreement) in cloud manufacturing industries. A robotic verification system of the certificate has been developed by Malsa et al. [36], where a blockchain network safely and immutably stores certificate information, and smart contract can check and verify the authenticity of the certificate. The application of blockchain with smart contract demonstrates that both of the technology works as a compatible dyad in which both blockchain and smart contract bolster each other's abilities and maximize their usability [72].

### B. Incremental Machine Learning

The predictive capability of a machine learning model depends on the feature it was trained on. The patterns and numbers (weights) that the model learns from a batch of data during training are prone to becoming obsolete in live applications where a billion instances of data are produced every second. A batch-learning-based ML model becomes incapacitated over time and requires a manual update which is computationally costly, laborious and impractical [9]. On top of that, batch-learning approaches lack the capability of incorporating the insights from new data instances in a deployed ecosystem [10]. Online machine learning is a technique where a trained and deployed machine learning model goes through continual upgrading. From a stream of data, the instances can be used one at a time (online learning) or a bunch of data (incremental learning) at one time. Incremental machine learning can aid the purpose of keeping the model updated over time, minimize the computation cost, and avoid manual updates [11]. In this machine learning approach, ML algorithms learn from a stream of data while the ML model is already deployed [10].

The growth of internet-based services and the shift of regular services into online platforms have generated a vast amount of data which will continue to grow further. A number of efforts have been seen in incremental machine learning over the past decade [37]. The use of incremental learning has been noticed in the medical sector [38], industrial process automation [39] and remotesensing [40].

Nilashi et al. used the advantages of incremental machine learning (incremental support vector machine) to build a predictor for Parkinson's disease [38]. Tian et al. have developed an incremental learning ensemble strategy (ILES), which incrementally learns from industrial sensor data to enhance the ability of soft sensors in an industrial environment [39]. Lin et al. used incremental learning to incrementally learn point cloud segmentation [40] and Fu et al. used the idea of incremental learning in end-to-end speech recognition [41]. However, most application of incremental learning is not class adaptive, and a solution to that problem was proposed by He et al. [42]. Their class adaptive approach also ensures that the model does not rely too much on new data (remembering its past experience).

### C. Collaborative Incremental Machine Learning Using Blockchain and Smart Contract

Data-driven technologies like machine learning often suffer because of data unavailability, knowledge centralization, and a monopolistic grasp of data [5]. Sophisticated data generated by finance or medical sectors are often kept private, and a few people have the opportunity to explore those data to extract new information and formulate new methods [4]. Extreme lack of trust among organizations, ensuring data privacy, maintaining data security and competitive market risk of data leakage are the reasons that these data are not shared [43]. Researchers have inspected the feasibility of decentralized ledger technology like blockchain aided by smart contracts as a solution to these problems of collaboration among organizations, privacy preservation, data security and breaking the centralization of both data and knowledge.

Research efforts have been seen to fuse blockchain and machine learning [14, 44, 45], where machine learning extracts patterns from data to generate predictions, while on the other hand, blockchain ensures the legitimacy of these data and ensures secure storage of data. Future generation technologies and ideas have started to adopt blockchain and smart contract technologies in assistance with ML techniques [46]. Wang et al. propose a distributed framework for energy trading where blockchain is used to protect the privacy of edge users, and machine learning techniques are used to generate accurate load and price prediction [47]. The blockchain-based microgrid fault identification method proposed by Liu et al. utilizes regional layering of power grids where machine learning algorithms are used for the timely identification of faults in power supply [48].

The study of Kim et al. [49] investigates the case of security, privacy and efficiency of machine learning integrated blockchain systems, and their system shows strong resilience against cyber-attacks. Vyas et al. show how the combination of blockchain and ML has been used in the medical sector [44]. The study by Liu et al. shows how a resource-efficient, scalable, secure and privacy-preserving blockchain with integration to intelligent machine learning techniques has been used in communication and networking systems [50]. Although blockchain ensures data security and legitimacy, they reported that private medical data is not shared, and due to that, machine learning research is hindered [44]. Solving these issues, many researchers have used blockchain to ensure differential privacy-preserving solutions in which the data identity is protected via cryptographic hash [45].

The use of machine learning techniques has shown promising performance in fraud detection [51]. As privacy-preservation and data security of blockchain-aided ML systems thrived, it opened the door to safe collaboration within the organization where no organization is at risk. Even the risk posed by insider intruders has been studied so that collaboration within organizations is not at risk from the node itself inside a blockchain network [52]. Tsuyoshi Ide showed a collaborative anomaly detection approach on a blockchain platform where blockchain has been used as a collaboration platform rather than merely a decentralized ledger storage system [53]. Chao Yang proposes a blockchain-based outlier clustering mechanism that utilizes the kernel density estimation technique and decomposition technique to cluster the incremental outlier features [54]. However, none of these studies explores the case of privacy preservation, task-specific incentive mechanism and developing efficient mining mechanism.

Justin D. Harris and Bo Waggoner have proposed a notable study on this topic [14]. Their work proposes a blockchain-based collaborative machine learning platform to build a dataset collaboratively, and they host their model using smart contract technology. Their work demonstrates a baseline implementation and suggests future work on privacy, using different ML models, devising new incentive mechanisms, etc. [14]. Inspired by the work of Harris et al., a trust-based collaborative movie recommendation filter has been developed by Yeh et al. [55]. All of their operations are done on-chain, and no performance analysis has been shown; on top of that, the incentive mechanism concentrated on collaborative contribution needs further attention [55]. Adaptive incentive mechanism is nasessary so that fair incentive is disbursed among the participants. On top of that, privary is a considerable issue in distributed learning platform in fintech. Considering the shortcomings and limitations, this study proposes a more unified approach leveraging the best of blockchain and smart contracts to build reliable machine learning models collaboratively while preserving privacy and increasing mining efficiency.

## III. Machine Learning Model Selection

There are two stages to this study. In the first stage, thorough experimentation is conducted with five machine learning algorithms to select the best model for deployment in the live environment. In the second stage, the necessary implementation of blockchain and smart contract is done to accommodate the ML algorithm to incrementally learn inside the blockchain-aided ecosystem. In this section, the experiments with five machine learning algorithms and their results have been shown to describe how the final model was selected.

### A. Machine Learning Models

We used the Scikit-Learn library for machine learning experimentation, which is open-source. This library allows direct API calls to machine learning models. Online incremental learning, while deployed, often becomes susceptible to catastrophic interference [8], which is when an ML model abruptly forgets its past experience and adapts the knowledge of a newly-fed dataset completely. Apart from training a model using the fit() method, Scikit-Learn also provides the functionality to train a model using the partial_fit() method incrementally. Currently, this library has five classification algorithms implemented. Our experimentation comprises these five algorithms. The five algorithms are Bernoulli naïve bayes, multinomial naïve bayes, passive aggressive classifier, stochastic gradient descent and perceptron.

The Bayesian algorithm is the set of all machine learning models which follows the probabilistic theorem of Thomas Bayes [56]. Bayesian probability calculation comprehends predicting the likelihood of an event based on some prior knowledge of known (labeled) data. The probabilistic Bayesian function considers the presumption that a feature under a class does not depend on the other features of that class, considering each feature to be class independent [56]. We used the Bernoulli Naive Bayes and the Multinomial Naive Bayes among the naive Bayes algorithms. The only difference between these two algorithms is the Bernoulli algorithm treats the feature as binary (0-No, 1-Yes) and the Multinomial treats the feature as a vector representation based on their occurrence, allowing the algorithm to work with multi-class problems. Passive Aggressive is another family of algorithms that can be used as a classifier which is a popular incremental online learning algorithm [57].

The passive-aggressive classifier responds as passive for correct classification and aggressively responds to misclassifications. For a correct classification, the algorithm yields a 0 value (passive), and for misclassification, the value is greater than zero (aggressive). Stochastic gradient descent is an iterative algorithm, and the main objective of this algorithm is to minimize the cost at each iteration [58]. After each learning iteration, the loss is calculated by comparing the prediction with the actual label. A minimizing formula then updates the value of the hyperparameters so that the error is minimized and the formula reaches the minima. Perceptron is the unit variant of a neural network having only one neuron. For a set of inputs ($x_1$, $x_2$, $x_3$, ....., $x_n$) and bias b, the output function Y tries to learn the weight matrix $\mathcal{W}$ [59]. The learning function repetitively runs till n times and opts to learn the weight matrix and bias value to minimize the error. The output is one if $w_i \cdot x_i > 0$ and set to 0 otherwise. The machine learning model for deployment is chosen based on the experiments done with these five machine learning algorithms: Bernoulli and Multinomial Naive Bayes, Passive Aggressive Classifier, Stochastic Gradient Descent and Perceptron.

### B. Dataset Description

A tremendous number of payments are done online on e-commerce platforms creating a vast amount of data. To represent this high-volume transaction data produced during digital payments, we have selected the "PaySim Synthetic Financial Datasets for Fraud Detection" [60], which is publicly available on Kaggle. This dataset is synthetically generated from an original dataset provided by a multinational fintech company that is successfully operating in fourteen countries. Analysis of their result shows that the synthetic dataset precisely resembles the original dataset [60]. The dataset contains 6.3 million instances of digital money transactions having a total of eleven features. The definition of each feature is as follows: ***1 ) step*** - step is a representation of real-time simulation. One step means one hour in real-time. The dataset is an assemblage of 744 steps taken over 30 days; ***2) type*** - the type of transaction (five types CASH-IN, CASH-OUT, DEBIT, PAYMENT and TRANSFER); ***3) amount*** - the amount of money transacted; ***4) nameOrig*** - identification credential of the sender; ***5) nameDest*** - identification credential of the receiver; ***6) oldbalanceOrg*** - account balance of sender before a transaction; ***7) newbalanceOrig*** - account balance of sender after a transaction; ***8) oldbalan*****ceDest -** account balance of receiver before a transaction; ***9) newbalanceDest*** - account balance of receiver after a transaction; ***10) isFlaggedFraud*** - this feature is a flag for transaction over 200000 in a single attempt; ***11) isFraud*** - this feature shows if the transaction is fraud or not and this is also the target variable.

### C. Feature Selection

Among the 6.3 million data instances, only 8213 instances were found to be fraudulent, which is roughly 0.0012% of the total data. On the other hand, the amount of money lost due to fraudulent transactions is roughly 0.010% of the total amount of money transacted. This implies a high amount of data imbalance among the data per class of the targeted attribute. Although our target attribute is a binary class (0-non-fraud, 1-fraud), there are five types of transactions infused into the whole dataset. It was found that only "TRANSFER" and "CASH-OUT" are the transaction types where the fraud cases happened. It is simply aboveboard that no fraudster would deposit money into a victim's account. Therefore, we discarded the other three classes and only took the instances of transfer and cash-out into account. It was also found that there was no transaction of more than 200000, which is why we also discarded the isFlaggedFraud feature. Finally, 2.6 million data was selected for machine learning experimentation. This final dataset has been entitled "experiment set," and this naming convention has been followed in the rest of the document.

### D. Evaluation Metrics

For an imbalanced data scenario, classification is a sophisticated task as only predicting the majority class will always result in high classification accuracy. To evaluate whether a classification algorithm has truly learned or not, a combination of different metrics is used alongside the confusion matrix [61]. The metrics we used in the study are illustrated in equation 1-6.

$$Accuracy = \frac{TP + TN}{TP + FP + TN + FN} \quad (1)$$

$$Sensitivity\ (TPR) = \frac{TP}{TP + FN} \quad (2)$$

$$Precision = \frac{TP}{TP + FP} \quad (3)$$

$$Recall = \frac{TP}{TP + TN} \quad (4)$$

$$Fbeta = (1 + \beta^2)\ \frac{precision * recall}{precision * \beta^2 + recall} \quad (5)$$

$$False\ Negative\ Rate = \frac{FN}{FN + TP} \quad (6)$$

In the equations above, TP, TN, FP, FN represents true positive, true negative, false positive and false negative, respectively, which is obtained from the confusion matrix. We used a modified f-score measure which is an f-score with a $\beta$ factor. The value of the $\beta$ factor makes the f-measure biased towards either precision or recall. In equation 5, the value of $\beta$ greater than one makes the Fbeta measure biased towards recall and the beta value less than one makes the Fbeta measure biased to precision. We used the recall-biased Fbeta measure as we want a deficient number of false negatives.

### *E. Experimentation & Selection of ML Models*

The machine learning experimentation starts with dealing with the high imbalance in the data. As mentioned in section 2.2, after the feature selection, a total of 2.6 million data were ready to experiment. But this severe skew resulting in high-class imbalance is problematic for machine learning algorithms to learn from [62] and often results in unsatisfactory performance. The fraud class in our dataset is the minority class, and we chose the oversampling strategy to balance the data first. Oversampling is prone to overfitting [62], and the issue of overfitting has also been investigated in our study.

To balance the majority and minority classes, we used an upsampling algorithm known as SMOTE (Synthetic Minority Oversampling Technique) algorithm [63]. SMOTE algorithm first determines the degree of imbalance ($d$) present in the dataset by capturing the amount of majority ($\mathcal{N}^-$) and minority ($\mathcal{N}^+$) classes. Another parameter, $\beta$ is determined by using $\beta = \frac{\mathcal{N}^+}{\mathcal{N}^-}$ which determines the proportion of majority to minority. Then the amount of minority data to be generated is determined by data = $\beta$ ($\mathcal{N}^-$ - $\mathcal{N}^+$). At the final stage, the KNN algorithm is used to generate projections of a randomly picked majority class and its neighbors. The newly added data is then included in the minority set of data.

The five machine learning algorithms were initially trained and tested using 67% of the experiment set (D1 in figure 2). The 67% of data (1742000 data instances) were split 80-to-20 percent for training and testing, which counts to 1393600 instances for training and 348400 instances for testing the initial model. We applied the SMOTE algorithm in both the training and testing set, producing 3473740 instances for training and 1112650 instances for testing. The rest of the 33% of the dataset (D2 and D3 in figure 2) is unattended at this stage and is preserved for testing the incremental learning phase inside the blockchain network. The metrics of the five models for the initial training and testing are shown in table 1.

Parameter optimization has been done for all the models using the GridSearchCV [64] method. For passive-aggressive, the value of c (maximum step size) was found to be 0.7%, tol (stopping criterion) to be 0.001% and maximum iteration to be 1000. From table 1, we can see that the passive-aggressive classifier (PAC) yielded the best performance among the five models. With a classification training and testing accuracy of 93.7% and 93.64%, respectively, PAC also has the best precision-recall and fβ scores. On the other hand, PAC has also scored the minimum false-negative rate. Perceptron model scored almost as good as PAC, but slightly better performance has been seen in the case of the PAC model. Contrary to that, Bayesian models (Bernoulli and Multinomial) moderately performed compared to PAC and Perceptron, while SGD showed an adequate performance. The confusion metric of the best-performing model (PAC) is shown in the figure below (figure 1).

**Figure 1. Passive-aggressive confusion matrix for initial training.**

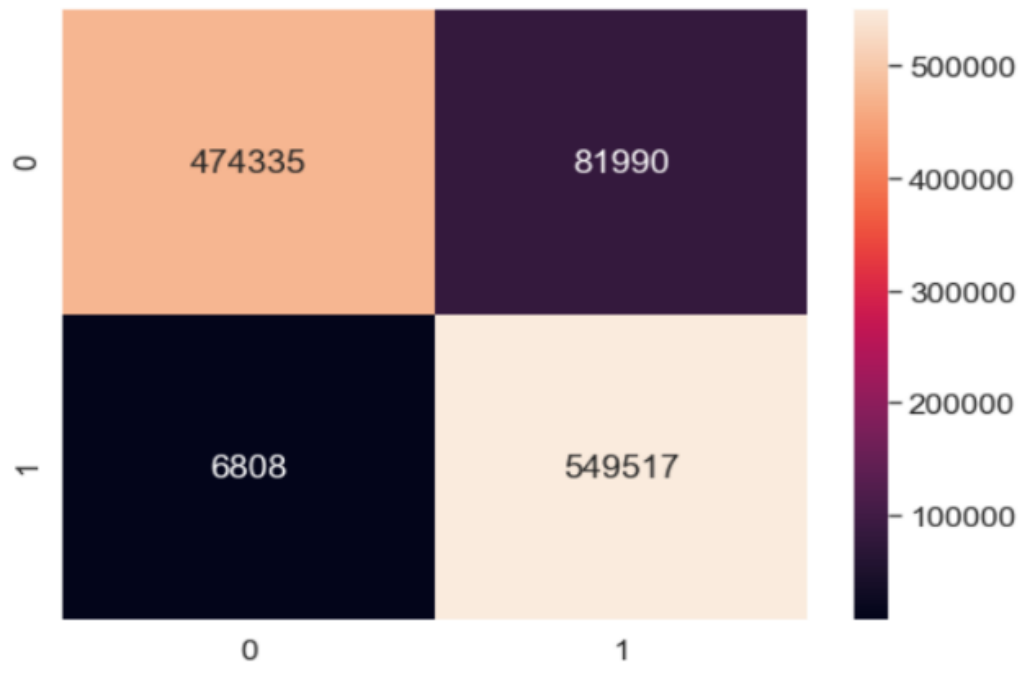

The initial model training is a batch training strategy where we select a fixed batch of instances for training and testing the models. However, it does not convey any solid indication that this performance will be reflected during the incremental learning phase. As PAC and Perceptron yielded an approximately similar result (with slightly better performance by PAC), we tested these two models for an incremental scenario before finally selecting a model which will be incorporated into our decentralized platform. From the 33% data kept separated (D2 in figure 2), we randomly picked 150000 data and made five datasets where each dataset containing 30000 instances. For a better understanding, the dataset preparation is again illustrated in figure 2, as the original data has been split into so many parts for different purposes.

We used these five datasets for testing the initially trained PAC and perceptron model to further inspect how they perform in an incremental learning scenario before deploying the model into our blockchain-based system. This incremental testing before deploying solidifies the selection of the machine learning model. Figure 2 illustrates that we have five experiment sets that we use to test the Perceptron and PAC algorithm. The metrics of Perceptron and PAC for these five experiment sets are shown in table 2. It is again clear that the passive-aggressive classifier performs slightly better than Perceptron. In the case of Perceptron, the metric (in table 2) improves in the second increment and then drops at the third increment, then again improves in the fourth and fifth increment. On the other hand, PAC shows a progressive growth of the metric for the five increments. Research also shows that passive-aggressive is a popularly used algorithm for incremental online machine learning tasks [57]. As both algorithms were tested with the same chunks of data, it is again established that a passive-aggressive classifier is a better choice for our specific purpose. Based on the experimentation described above, the model we chose for deploying is a passive-aggressive classifier into our system.

TABLE I
INITIAL MODEL TRAINING RESULTS

| ML Model | Training Acc | Testing Acc | Precision | | Recall | | $F\beta$-Score | | False Negative Rate |
|---|---|---|---|---|---|---|---|---|---|
| | | | 0 | 1 | 0 | 1 | 0 | 1 | |
| SGD | 85.58 | 86.55 | 0.99 | 0.78 | 0.72 | 0.99 | 0.83 | 0.87 | 0.059 |
| Perceptron | 92.56 | 92.63 | 0.94 | 0.91 | 0.91 | 0.92 | 0.93 | 0.93 | 0.058 |
| MNB | 80.86 | 80.73 | 0.84 | 0.78 | 0.76 | 0.86 | 0.80 | 0.82 | 0.14 |
| BNB | 72.70 | 72.15 | 0.69 | 0.77 | 0.82 | 0.62 | 0.75 | 0.69 | 0.38 |
| PAC | **93.7** | **93.64** | **0.96** | **0.92** | **0.94** | **0.96** | **0.93** | **0.94** | **0.041** |

TABLE II
INCREMENTAL LEARNING COMPARISON BETWEEN PASSIVE-AGGRESSIVE AND PERCEPTRON.

| Increment No | Training Acc | Testing Acc | Precision | | Recall | | $F\beta$-Score | | False Negative Rate |
|---|---|---|---|---|---|---|---|---|---|
| | | | **0** | **1** | **0** | **1** | **0** | **1** | |
| **Metrics of Perceptron Algorithm for Five Increments** | | | | | | | | | |
| 1 | 90.66 | 90.17 | 0.90 | 0.91 | 0.92 | 0.88 | 0.91 | 0.92 | 0.079 |
| 2 | 93.23 | 91.55 | 0.91 | 0.92 | 0.92 | 0.90 | 0.92 | 0.92 | 0.013 |
| 3 | 92.66 | 92.68 | 0.92 | 0.87 | 0.85 | 0.91 | 0.92 | 0.93 | 0.010 |
| 4 | 94.89 | 93.81 | 0.93 | 0.94 | 0.93 | 0.92 | 0.93 | 0.93 | 0.023 |
| 5 | 94.99 | 95.87 | 0.95 | 0.94 | 0.94 | 0.95 | 0.96 | 0.96 | 0.049 |
| **Metrics of Passive Aggressive Classifier for Five Increments** | | | | | | | | | |
| 1 | 92.48 | 92.50 | 0.91 | 0.91 | 0.92 | 0.92 | 0.93 | 0.92 | 0.075 |
| 2 | 92.86 | 92.56 | 0.92 | 0.91 | 0.93 | 0.92 | 0.93 | 0.91 | 0.056 |
| 3 | 94.40 | 93.38 | 0.93 | 0.94 | 0.93 | 0.93 | 0.93 | 0.94 | 0.049 |
| 4 | 94.20 | 96.42 | 0.94 | 0.94 | 0.97 | 0.97 | 0.95 | 0.96 | 0.026 |
| 5 | **96.83** | **96.86** | **0.95** | **0.96** | **0.95** | **0.96** | **0.96** | **0.96** | **0.010** |

**Figure 2. Data split for initial batch training and incremental learning experiments.**

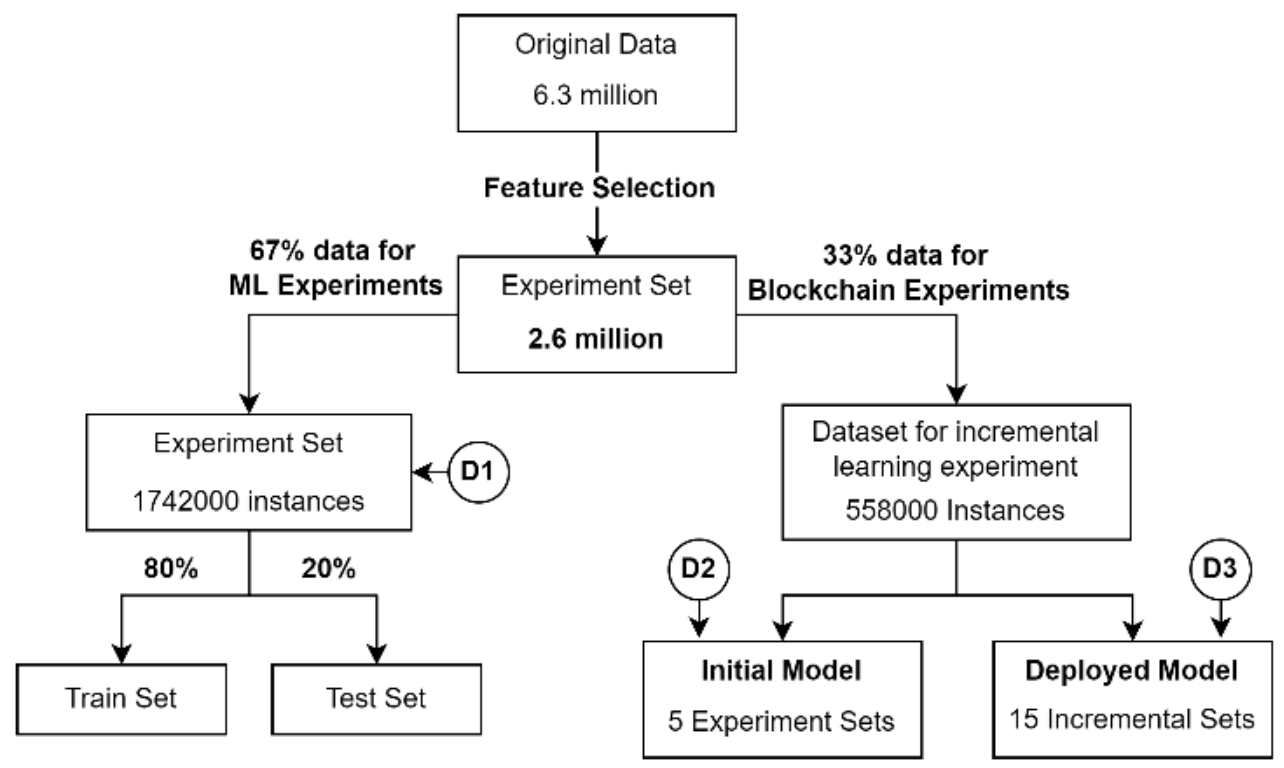

## IV. System Design of Blockchain and Smart Contract Based Decentralized Application

The sophisticated nature of financial transaction data is deteriorating for open and collaborative research [3]. Due to privacy issues, market competition and confidentiality, transaction data are not shared publicly. And on the other hand, financial fraud causes an enormous amount of loss to the whole industry. Our system implements a smart contract-aided blockchain system where organizations can collaboratively contribute to a machine-learning algorithm to learn incrementally while preserving privacy and confidentiality. This section delves into a detailed description of our proposed system and its core architecture. The core components of our system are shown in figure 3. The detailed system design, the definition of the actors that are connected to our system, and a comprehensive workflow of our proposed architecture has been shown in the following subsections.

### A. System Actors

There are three types of actors within our system. Regulators are the primary authority to initially deploy the machine learning model into the system. A regulator will also deploy the smart contract used in the system. After the regulator initiates the system, it is automated afterward. The second type of actors are contributors who are the e-commerce organizations. Contributors are the organizations who will contribute to the machine learning model by providing genuine data. As data privacy is preserved in our system, the contributors can dauntlessly contribute and get an incentive for successfully updating the machine learning model. The most recent updated machine learning model is openly available. A user can use this model on a per-query basis without any charge.

### B. Overview of Proposed Architecture

The system architecture is comprised of three layers, the application layer, the off-chain machine learning layer and the blockchain layer. The application layer is the user interface where participants (contributor, user) can register, use the machine learning model and contribute with their organization's transaction data to improve the ML model. A decentralized application connects the actor to our blockchain network in the application layer. The selected ML model (Passive-aggressive classifier) is deployed in the blockchain at the inception of the network (the genesis block of a blockchain). When a contributor or user node joins the network, they can use the ML model to perform queries (use ML model) regardless of their role within the system. However, the participant node must be a contributor to commit data to the system for partially training the model. After a contributor submits a new instance of the dataset, the request via the distributed application is sent to the interface (API). The interface then passes the data to the ML layer of our application.

**Figure 3. The core components of the system.**

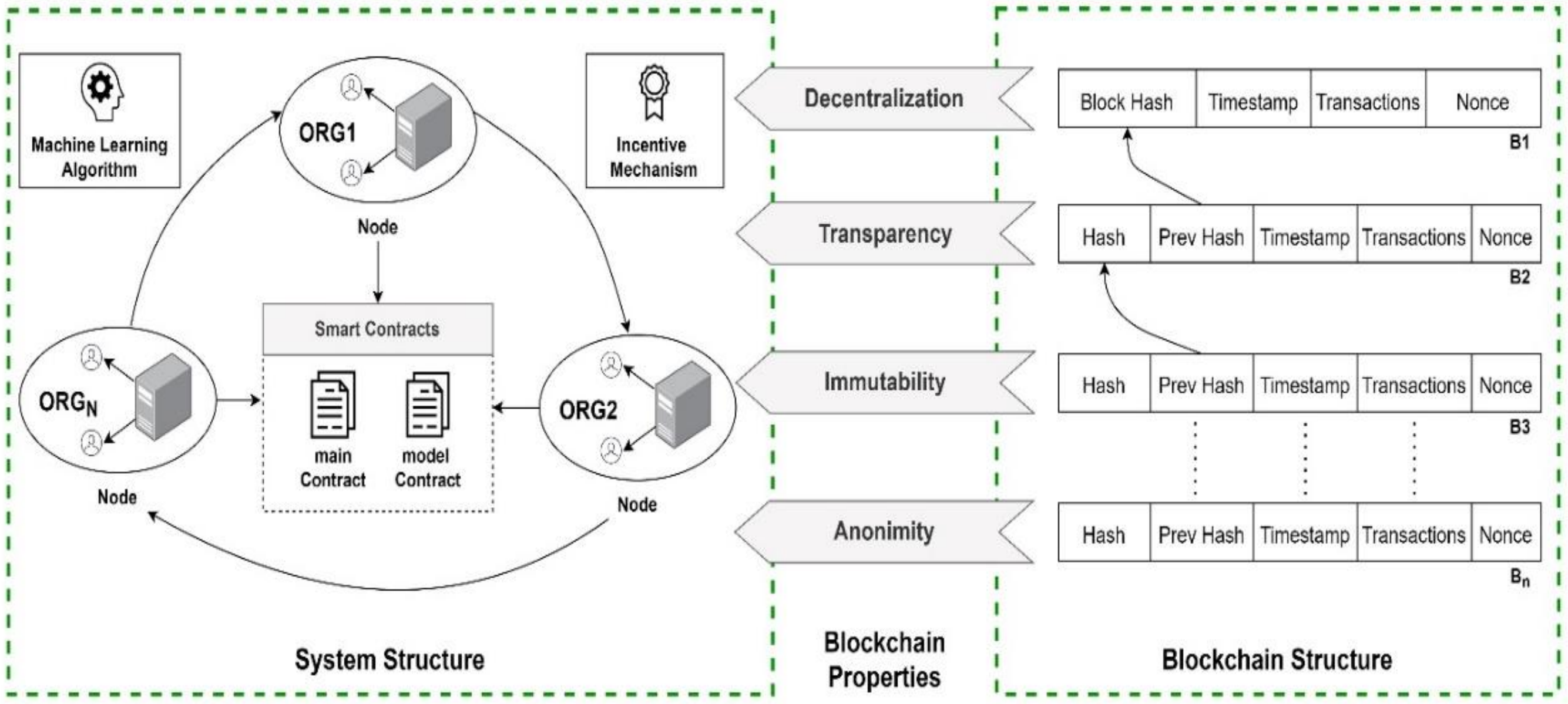

Figure 4 illustrates that the ML layer is composed of all machine learning requirements starting from data preparation to partially training the model. The data passed by the application layer first goes through a data preparation filter where the dataset is checked for trivial cases like data shape, necessary feature presence, null values, etc. Resembling the initial dataset, these incoming datasets are also expected to be imbalanced, and at this stage, SMOTE algorithm is applied to diminish the class imbalance. Then the balanced dataset is split into train and test sets. The train set is used to partially train the current ML model, and the test set is used to evaluate the model's performance after the partial training. The metrics are off-chain and awaiting inclusion in the blockchain network at this stage.

Metric and results from the ML layer are sent to the main contract first. The main contract analyzes the current metrics compared to the previous ones and decides (using algorithm 1) whether it is an improvement. Then the decision is sent to the model contract. If the decision indicates improvement, a new block with the model metrics and results is created and broadcasted to the network to include it in the main chain. The model contract also updates the current ML model in case of improvement. The model contract also calculates the incentive (using equation 10) and adds it to the address (organization) from which the data was provided. The learning history (metrics) of the ML model is also stored in the blockchain network. The cryptographic hash of the updated model is stored in the blockchain and this updated model will be used by the ML layer and the application layer in further steps. The application layer shows the live update of the current blockchain, ML model, transactions, etc.

**Figure 4.** **Detailed system architecture.**

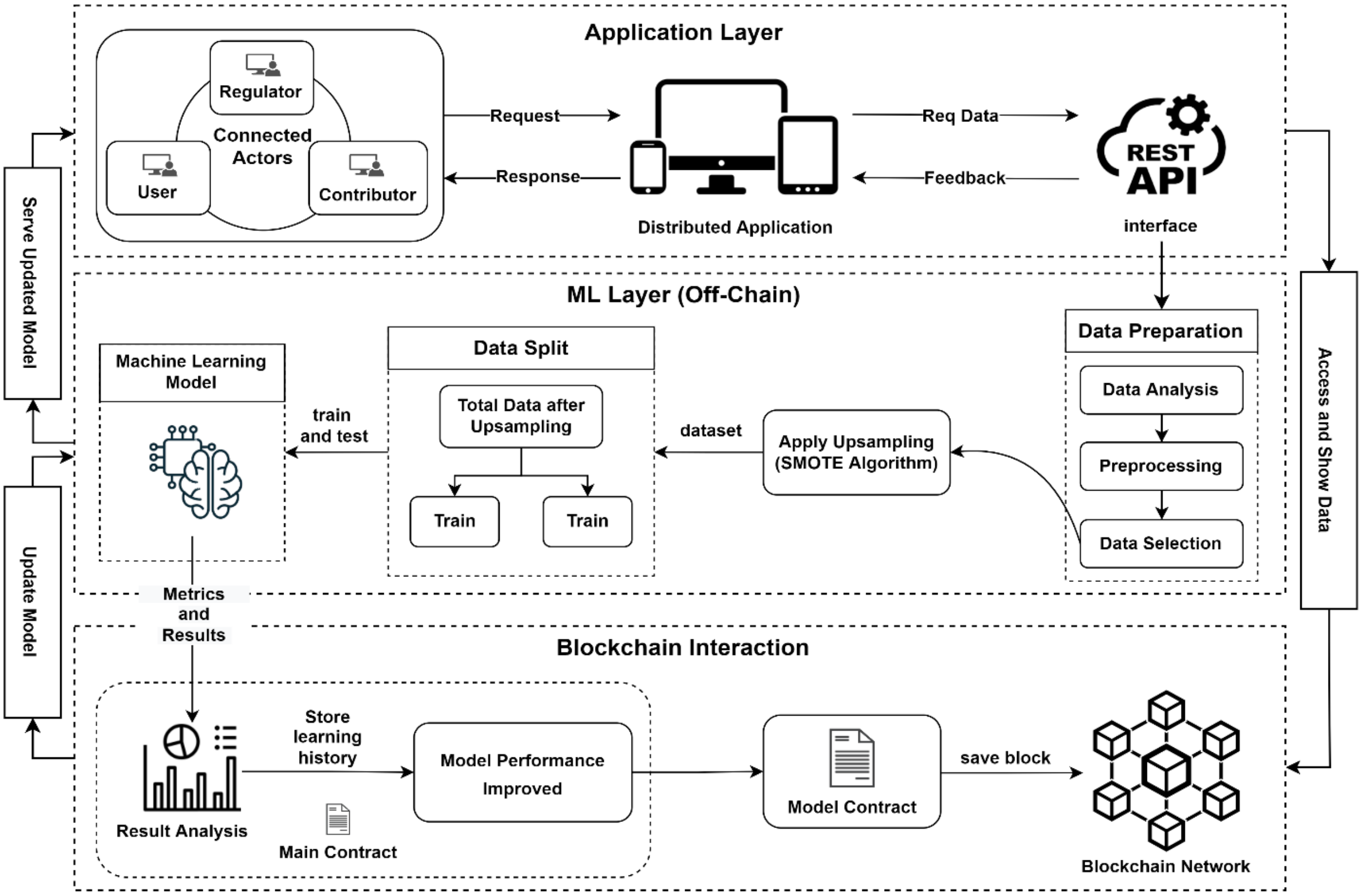

## V. Setting up The Test Environment

### A. Developing the Blockchain

This section contains an intricate discussion of our system implementation. As discussed in section 4.3, the system has three layers to be implemented. Linux 20.04 LTS was used as the operating system to develop the system, while the CPU is Intel(R) Core(TM) i7-8565U 1.80GHz having 4 Core(s), 8 Logical Processor(s) and a total RAM of 16 GB. We used the VSCode IDE for coding the system. Python language has been used to implement the architecture, including the machine learning, blockchain, smart contract and experimentation phases. The python flask library has been combined with HTML5, CSS, and Bootstrap to develop the web application. The flask library also provides supporting privileges like data interfacing, machine learning integration and data analysis. Python version 3.9.6 and flask version 2.0.2 was used during the coding implementation. The blockchain and smart contract we used were also written in the python programming language. The blockchain was implemented as a python class, and the Blockchain class inherits another class named the Block class. The Block class defines the structure of a single block. The content of a single block is the block number, block hash, previous block hash, nonce, timestamp and data.

Considering each entry as a transaction T, a single block in our system represents a collection of entries to the block. A block $\mathbb{B}$ containing n entries is defined as the following equation. (Equation 7)

$$\mathbb{B} = (T_1, T_2, T_3, T_3, \dots \dots \dots T_n) \tag{7}$$

When a new batch of data improves the previous model, only then a new block is created and the node itself will first find the 'nonce' such that the following equation satisfies.

$$\mathcal{H}\left[\mathcal{H}(\mathbb{B}^{i-1}) \oplus R(\mathbb{B}) \oplus T(t) \oplus d \oplus nonce\right] \leq \text{target} \tag{8}$$

Where, $\mathcal{H}$ is the hashing function, $\mathcal{H}(\mathbb{B}^{i-1})$ represents the hash of the previous block, $R(\mathbb{B})$ stands for the Merkle root, T(t) denotes the timestamp at time t, and d represents the difficulty level. The concatenation ($\oplus$) of these terms must generate a hash less than or equal to the previously known target. The proof-of-work consensus [65] has been used where the other nodes verify the legitimacy of a newly generated block before adding it to the main chain. The miner node provides the timestamp($t^c$) and a nonce ($n^c$) while broadcasting the new block, and the other block can effortlessly verify the hash using the following equation.

$$\mathcal{H}\left[\mathcal{H}(\mathbb{B}^{i-1}) \oplus R(\mathbb{B}) \oplus t^c \oplus d \oplus n^c\right] \tag{9}$$

The data inside a block in our system represents the machine learning model's learning history (metric in each increment). The data has been represented in JSON format throughout the whole system. Every cryptographic hashing in our system uses the SHA256 hash function, which generates a 64-bit hexadecimal hash. The Blockchain class implements five functions to add, mine or remove a block to the blockchain, check the integrity of the blockchain, and return the current chain to any caller. The mining and addition function implementation is shown in figure 5.

### A. Developing Smart Contracts

Two smart contracts were developed to support the automated and secure maintenance of the machine learning model inside our blockchain network. The smart contracts are constructed by combining attributes, events, functions, modifiers, etc. The first contract, namely, the main contract, works as a controller working between the user application and the system. This contract receives the model metrics after each partial training and then analyzes them to decide whether to update the ML model. If the decision is positive, only then the metrics and decision are sent to the model contract. The model contract updates the current hash of the ML model to the updated model's hash. And finally, the model contract calculates incentive and sends the amount to the appropriate contributor account. The attribute, function, event and modifiers used in our contracts have been shown in table 3.

**Figure 5. Code for mining and adding a block to the blockchain network.**

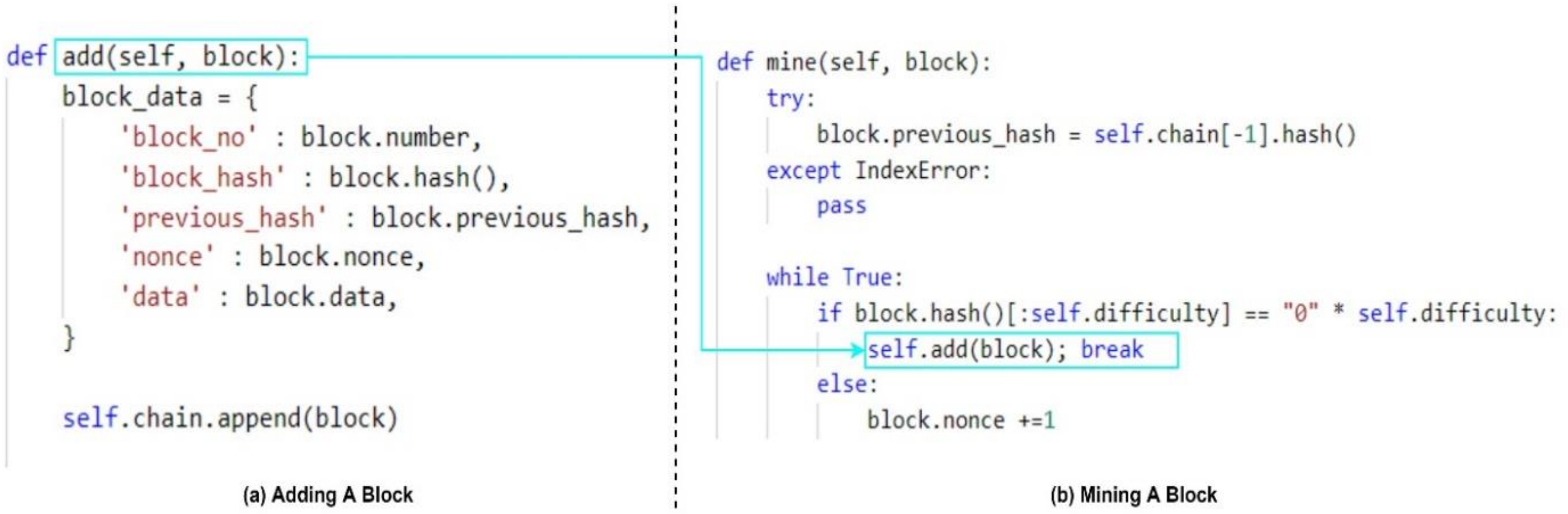

TABLE III
SMART CONTRACTS AND THEIR COMPONENTS WITH DESCRIPTION

| Contract Component | Type | Description |
|---|---|---|
| **Components of the main contract & their description** | | |
| Best precision, recall, $f\beta$ and false-negative scores | Variables | Stores all the current best metric scores for calculating the model hash, incentive etc. during and model update. |
| Only Regulator<br>Only Contributor<br>Only User | Modifier | Controls the access of the users based on their role; for example, only a contributor node can add data to the system while a user node can only access the machine learning model. |
| Compare Result | Function | Compares the current metrics with the previous best model's result. If the current model yielded better results returns a positive response. |
| Calculate model hash | Function | For a model update, this function calculates the model hash by taking in its metrics, e.g., precision, recall, Fbeta and false-negative rate. |
| **Components of the model contract & their description** | | |
| Base incentive | Variable | When a model is updated, a special incentive is given which is much greater in amount, but for cases where good quality data couldn't update the model, the base incentive is given. |
| Current model hash | Variable | Stores the hash of the currently available best model. |
| Calculate incentive | Function | Calculates the incentive considering the improvement of the model by the recent dataset provided by a contributor. |
| Update model | Function | Update a better-performing model replacing the previous one. |
| Updated model | Event | Emit an event in the network notifying that there is a newly updated model available with better performance. The model hash and metric will also be emitted while calling this event. |

The combination of these events, functions and modifiers provides the organizations to safely collaborate as well as earn incentives. A successful contribution resulting in a better result than the previous best results will receive an incentive based on the following formula (equation -10).

$$\mathbb{I} = P_i + \frac{[( R_i - R_B)*(P_i - P_B)*(F\beta_i - F\beta_B)]^2}{\gamma*(FNR_i - FNR_B)} \quad (10)$$

In equation 10, P represents the current price of the model (initially set to 100 in our system), and the notation $R$, $P$ and $F\beta$ stand for recall, precision and fbeta-score, respectively. Subscripted 'i' represents the current iteration number of model update, and B represents the previous best value. The $\gamma$ factor in the denominator ascertains that the incentive function always produces a value greater than the previous incentive.

As the model improves, the difficulty of surpassing the result also increases. So, whichever contributor could improve a model, should be paid more than the previous contributor. To achieve such a mechanism, the value of $P_i$ is updated as $P_i = \mathbb{I} + P_i$ after an incentive is disbursed. The value of $\gamma$ is initially set to 0.001 (empirically decided), and after every update, the value is increased by adding 0.002 so that the next incentive produces a greater amount using equation 10. The incentive disbursement process is governed by the smart contracts automatically. When an organization contributes new data that successfully updates a model, the incentive value is added to the organization's address.

## VI. System Testing & Result Analysis

The preceding sections have described our machine learning model selection, architecture and implementation details. This section contains an out-an-out testing of our model from blockchain, smart contract, and machine learning perspectives.

### A. System Testing

At the very beginning of our system, a contributor initiates the system by training the first model for deployment. The model has already been chosen based on some initial experiments described in section 3. The initial model training functionality is shown in figure 6.

**Figure 6.** **A regulator training the initial model to be deployed in the system.**

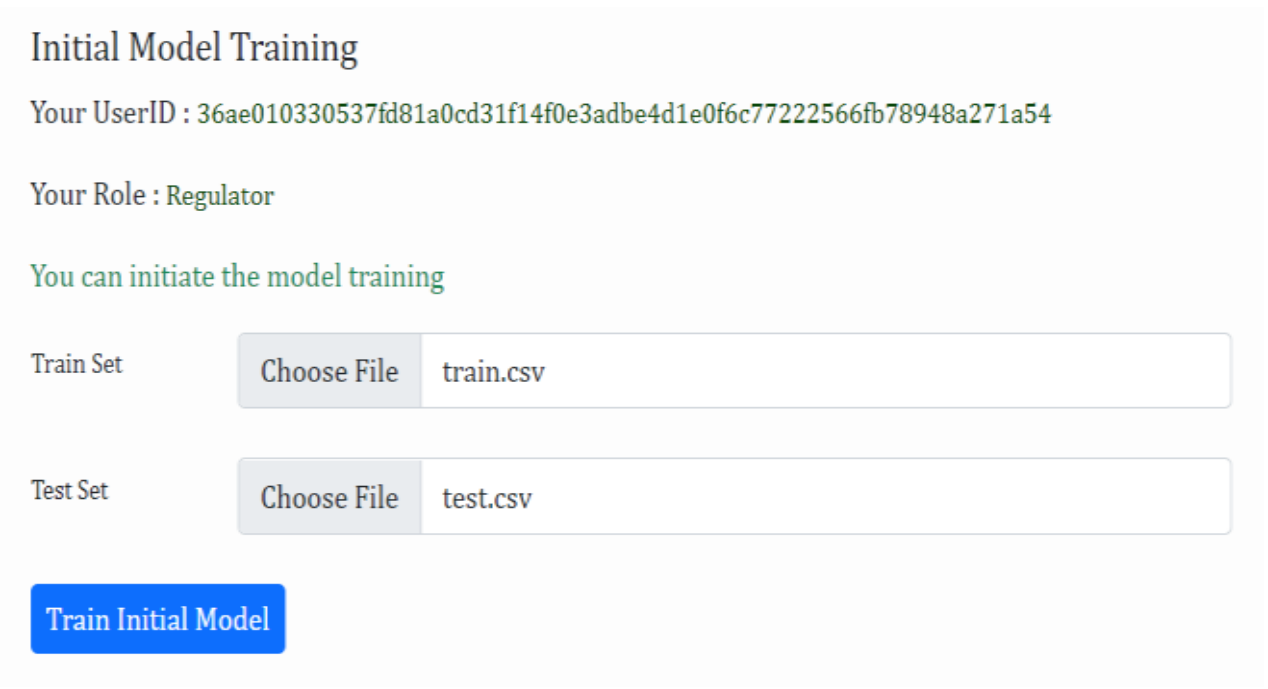

After the initial model is dispatched within the system, the contributors can now contribute to the model, and the users have this model available to perform a query. Figure 7 shows the contribution by a contributor node to the existing ML model.

**Figure 7.** **A contributor node contributing to the existing ML model.**

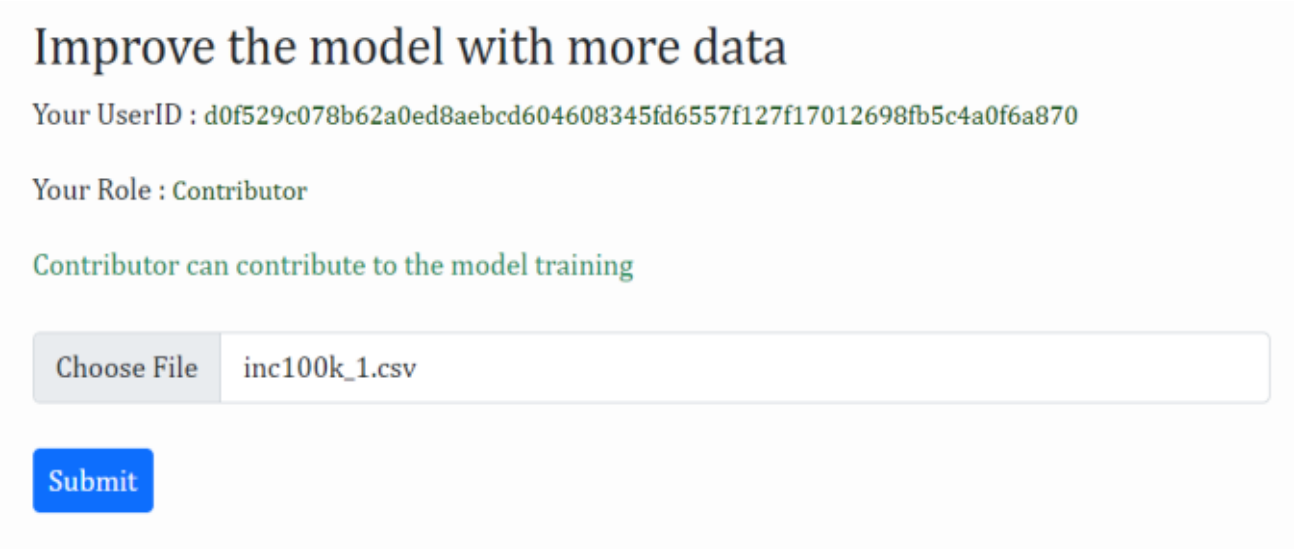

While partially training with new data (provided by contributors), the deployed model itself is unaffected. Instead, a copy of the deployed model is always kept for experimentation. The copy of the current best ML model is partially trained for a new dataset. And, if it results in a better false-negative rate while also keeping other metrics (precision, recall, Fbeta) in the average range, only then is the model updated. The fundamental reason is to build the model robust to detect false negatives as the false-negative cases cause financial loss to an organization. The f-measure we used is already weighted to better recall, which solidifies the reason to choose Fbeta as a parameter to acknowledge before updating a model. The model update decision and the steps during a model update construct our mining criteria which are the collective steps before mining a new block. The mining criteria are described as the following algorithm.

**ALGORITHM 1: MINING CRITERIA.**

**Result:** Model update decision.

**compare_result (current_metric, previous_best_metrics):**

  **if** $FNR_{current}$ is better than $FNR_{best}$ **then**

    **if** $\left(\frac{\sum_{i=1}^{n} P_i}{n} \cong p_{best}\right)$ **and** $\left(\frac{\sum_{i=1}^{n} R_i}{n} \cong R_{best}\right)$ **and** $\left(\frac{\sum_{i=1}^{n} f\beta_i}{n} \cong f\beta_{best}\right)$ **then**

      **calculate_new_model_hash**(p, r, $f\beta$, fnr)

      **update_current_model(**new_hash**)**

      **emit event** regarding the model update

      **calculate_incentive**()

      **mine_block**()

  **else**

    skip

Where P and R stand for precision and recall, respectively. Following these conditions stated in algorithm 1, if a newly trained model outperforms the previous model's metric, a block of JSON file is created (a new block). The new block has the contents such as the current block number, hash of the new block and hash of the previous block, the nonce, and the model metrics. When a new block is created, an existing best model is replaced with a better ML model (newly trained model). The smart contracts calculate (using equation 10) and disburse the incentive to the address from which data has been uploaded (the organization). The new block is broadcasted to the network and eventually added to the main chain. Thus, a new block is added and stored in the blockchain network. Figure 8 shows the content of a block after a successful model update.

**Figure 8.** **Creation of a block after a model update.**

```
{
    "block_no": 8,
    "block_hash": "00000e565ba884aff27c7b4938dc6d991a345547dcbcf26d5b275114ae617dcf",
    "previous_hash": "00000480a3c397d69bcef6109b718811b4dfa8ca4570de8ca227d70c57af38dd",
    "nonce": 463442,
    "data": {
        "taining_accuracy": 0.9889713609892659,
        "testing_accuracy": 0.9751902504254801,
        "overall_accuracy": 0.9351902504254802,
        "precision": 0.9810447917216556,
        "recall": 0.9851902504254801,
        "f1score": 0.9349744590125882,
        "fbeta": 0.988417249023909,
        "true_positive_rate": 0.9927972283005105,
        "true_negative_rate": 0.8775832725504498,
        "false_positive_rate": 0.1224167274495502,
        "false_negative_rate": 0.007202771699489424,
        "hash": "c84aa72fea2c9ff3b2461d12b6931e77dcaa09aa4fa66789f292c715bee986ce"
    }
}
```

As depicted in figure 8, all the metrics along with the model hash has been stored inside the block. Figure 9 shows the free usage of the ML model for performing a query.

Figure 9. **Openly available ML model used by the user nodes.**

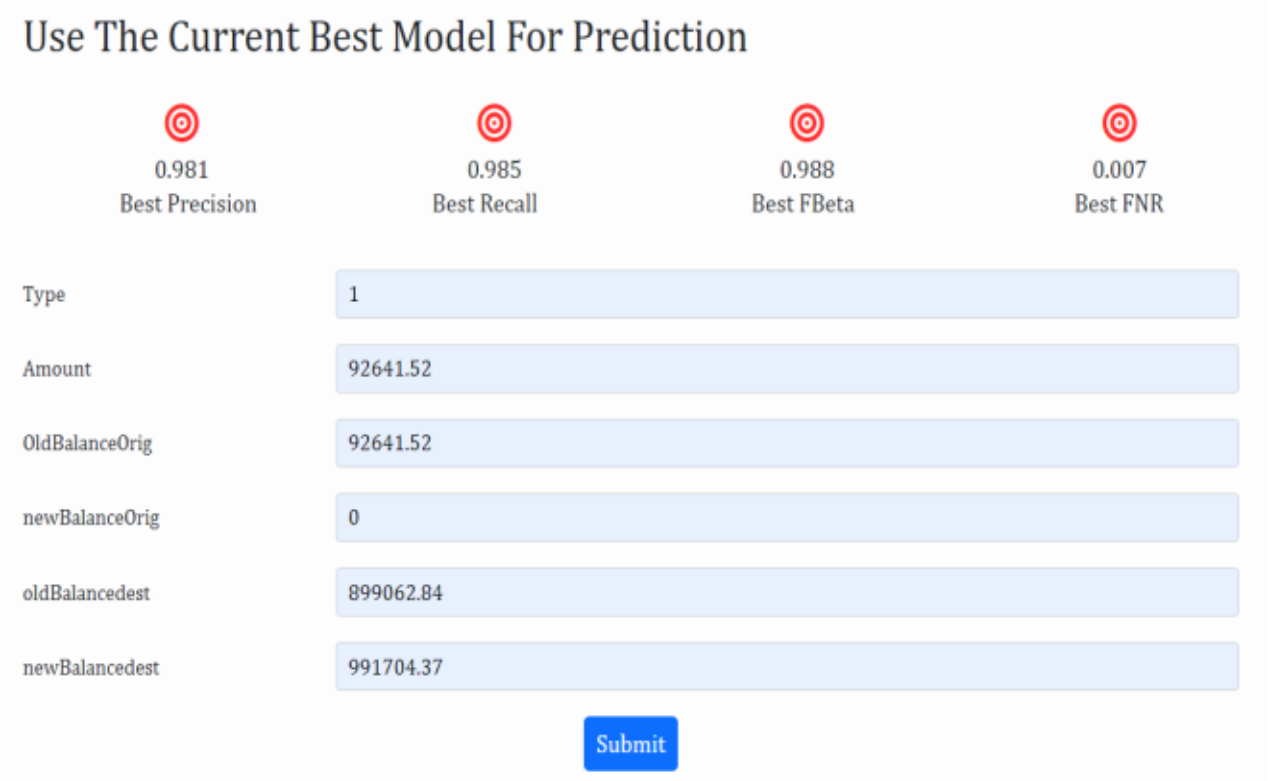

All these functionalities in our system are accessible to all kinds of users via the dashboard shown in figure 10. The vertically arranged card on the left is the counters. The total contribution counter shows contributors' total number of datasets uploaded to the system. The model update card shows how many times the model has been updated, outperforming its precedent model. The query served card shows how many queries have been served by the current ML model in a single day. We tested the system by providing 40 datasets of different volumes (described in section VI.B). The contribution and model update number often vary because it is not certain that a new dataset updates the model as well. As seen in figure 10, eight datasets were able to improve the models' performance, and hence, eight updates of the model have been recorded by the system. The overall steps from selecting the PAC model to incrementally updating the model in blockchain have been depicted in algorithm 2.

### *A. Mining time analysis*

Figure 6-9 demonstrates a successful run of our system functionalities. However, to record the data inside the blockchain, the block containing model update information needs to be mined, which creates latency. Time is crucial in real-world applications where convenience is among the top priorities. In a blockchain network, the latency is created by the difficulty level of the blockchain network [66]. As the difficulty increases, it becomes difficult for the miner node to generate the nonce and mine the block [60]. We examined the network under different volumes of data and under different difficulty levels. Five computer machines with varying computational power were used to analyze the mining time. The description of the machines is given in table 4.

Figure 10. **User interface of the application.**

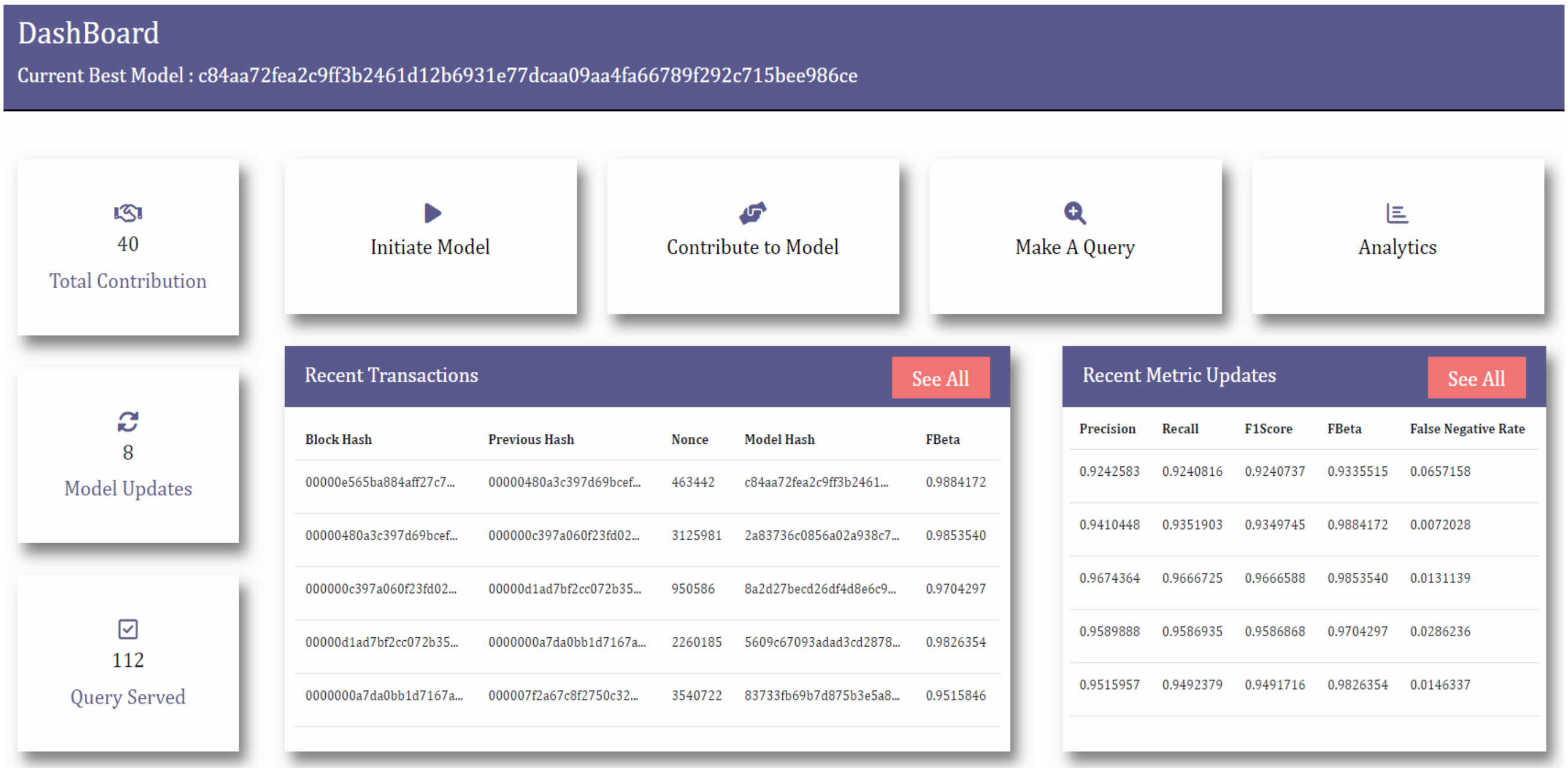

**ALGORITHM 2: PSEUDO STEPS OF THE SYSTEM: FROM MODEL SELECTION TO INCREMENTAL LEARNING IN BLOCKCHAIN**

**Result:**

- Selecting the suitable ML model to deploy in the system.
- Developing a privacy-preserving system for safe collaboration.
- Incrementally update the ML model and make it robust.
- Efficient mining criteria for optimized mining.
- Difficulty adaptive incentive mechanism.

01 **Pre-deploy ML experimentation:**

The best machine learning model is selected among five ML models (Bernoulli and Multinomial Naïve Bayes, Passive Aggressive Classifier, Stochastic Gradient Descent and Perceptron) using dataset D1 (figure 2). Passive Aggressive and Perceptron were found to be performing relatively well; hence, they were initially chosen for deployment, awaiting final selection.

02 **Pre-deploy incremental experimentation:**

The initial choice was made based on a batch-training experiment. Before deploying the model in blockchain, both the models (PAC and Perceptron) were tested incrementally five times. Using dataset D2 (figure 2). Passive-aggressive was found to be performing better than Perceptron. PAC model was chosen to be deployed in the blockchain.

03 **calculate_model_hash(model metric)**

The hash of the PAC model (selected is step 2) is calculated using the precision, recall, Fbeta and false-negative rate of that model.

04 **deploy_model(model hash):**

- The genesis block (first block of a blockchain) is created using the model hash along with the learning history (metric) from the regulator node.
- The hash of this model is updated as the current best model and the metric of this model is set as the global best metrics inside the system.
- The next iteration of the ML model is decided based on these global best metrics.

05 **partial_training(dataset file):**

When a contributor node uploads a dataset to contribute to the model, the partial training functionality steps are followed chronologically. The steps are as follows:

06 **data_preperation(dataset):**

Data filtration, shape matching, null value detection and feature matching.
Apply SMOTE algorithm to balance the data.
**return** balanced data

07 **partial_training(**balanced data**):**

- A copy of the current best model (from step 4) is partially trained.
- Then algorithm one is followed. Algorithm 1 only allows a model update given that the new data has improved the currently deployed model. In this step, the incentive is calculated based on the difficulty of updating the model (following equation-10) and the contributor is awarded by the calculated amount. Finally, a block gets created to store the learning history.
- All the calculations and data usage are done off-chain (on the node machine), securing the privacy of the organization.
- To simulate the incremental learning in blockchain, we used 15 incremental sets of 75k from dataset D3 (figure 2). Eight of them were successful to improved their previous best metrics.

08 **Model Usage:**

The model is openly available to the connected nodes. The organization themselves can call the API to use the model on the back-end of their website.

TABLE IV
COMPUTATION POWER OF FIVE MACHINES USED FOR MINING

| Machine No. | CPU | RAM | GPU |
|---|---|---|---|
| M1 | Core i5-6500U | 12 GB | Nvidia 1660 Super |
| M2 | Ryzen 5-3500G | 16 GB | Nvidia 1650 |
| M3 | Core i5-7500U | 8 GB | Nvidia 1060 |
| M4 | Core i7-7700HQ | 8 GB | Nvidia 1050Ti |
| M5 | Core i7-8565U | 16 GB | Nvidia 1050 Max-Q |

The initial model is trained with a relatively larger dataset (1.3 million). The average time taken to mine the initial model has been tested under blockchain difficulty level two to level five, which is depicted in figure 12.

**Figure 12. Mining time analysis of the initial model.**

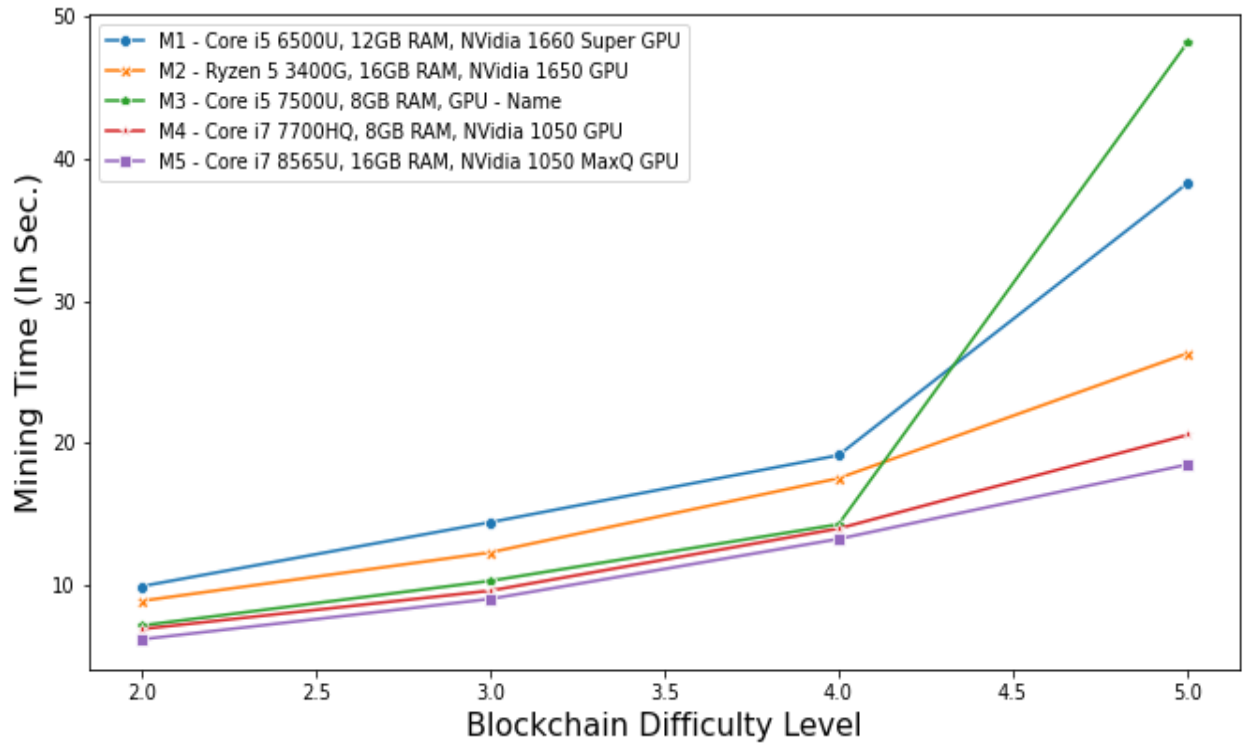

Figure 12 delineates the mining time of the initial ML model with a relatively larger dataset. The mining process takes as low as 2 seconds and as high as around 50 seconds to mine the block, depending on the difficulty level. On the other hand, for difficulty levels 2-4, the comparatively faster machines mine the block taking the least time. For difficulty level five, an exception to this trend is seen where M3, a comparatively more powerful machine than M1 and M2, took a longer average time to mine the initial model. After the deployment, the continual learning process starts, where the model is partially trained with relatively more minor datasets simulating the scenario where organizations will contribute their weekly/monthly data. During our initial experimentation, a fraction of the original dataset was kept separated (shown in figure 2) for testing the incremental learning of the model deployed in the blockchain network.

The experimentation starts with five datasets of 50 thousand, and the initial difficulty level is set to 2. All five machines were tested to determine the mining time trend in the network. At each step, the dataset size increased by 25 thousand, but the instances were randomly chosen from the dataset used for deployed training. Figure 12 delineates the mining time graph for increasing difficulty levels. The experiment ends with five datasets of 200 thousand at difficulty level 5. Figure 13 shows a similar trend during incremental model training. An increasing trend is discovered in the data for difficulty levels 2-4 and difficulty level 5 presents a much more difficult problem for the machine to mine the block, resulting in scattered mining time among machines with varying computational capacity.

For difficulty 2 and 50 thousand data shown in figure 13 (a), the mining time is below 50 milliseconds to the lowest and close to 800 milliseconds to the highest for 200 thousand datasets. For the same amount of data, the machines with comparatively high computation power (M5) are taking low mining time and the opposite is seen on a relatively less powerful machine (M1). On the other hand, as the volume of data is increasing, the mining time on a particular machine is also increased. For instance, in figure 13 (a), the purple color line plot shows the growing movement of the graph as the volume of data increases. The same trend is repeated in figure 13 (b) and 13 (c) for difficulty levels three and four, respectively. However, difficulty level 5 does not properly repeat the pattern in mining time. For instance, M1, which is a comparatively less powerful machine takes much less time (approximately 10 seconds) to mine the model which was partially trained with a dataset of 200 thousand instances.

### *B. Incremental Learning in Blockchain*

Eight of the 40 simulated datasets enhanced the model by outperforming its previous model in terms of metrics (shown in figure 10). One of the significant aspects of this study is to build a robust fraud detection ML model through collaboration. And, to bring robustness with a view to minimizing the financial loss to its lowest, our model focuses on controlling the recall to gain a much better Fbeta and false-negative rate. Figure 14 shows the confusion matrix for each of the eight increments. The figure shows a decreasing amount of false-negative at every increment. Otherwise narrated, only the model which has a much lower false-negative has been chosen to replace the previous model. Starting from a false-negative amount of 3922 for increment 1, through incremental learning, the model improved and at increment eight the amount of false negative was reduced down to 729 on the same amount of data. Figure 14 delineates that the amount of false-negative is decreasing with each increment which was the target to achieve in this study. Figure 15 shows every parameter update on each step of the machine learning model update.

**Figure 13.** Mining time graph for incremental learning under increasing difficulty and data volume.

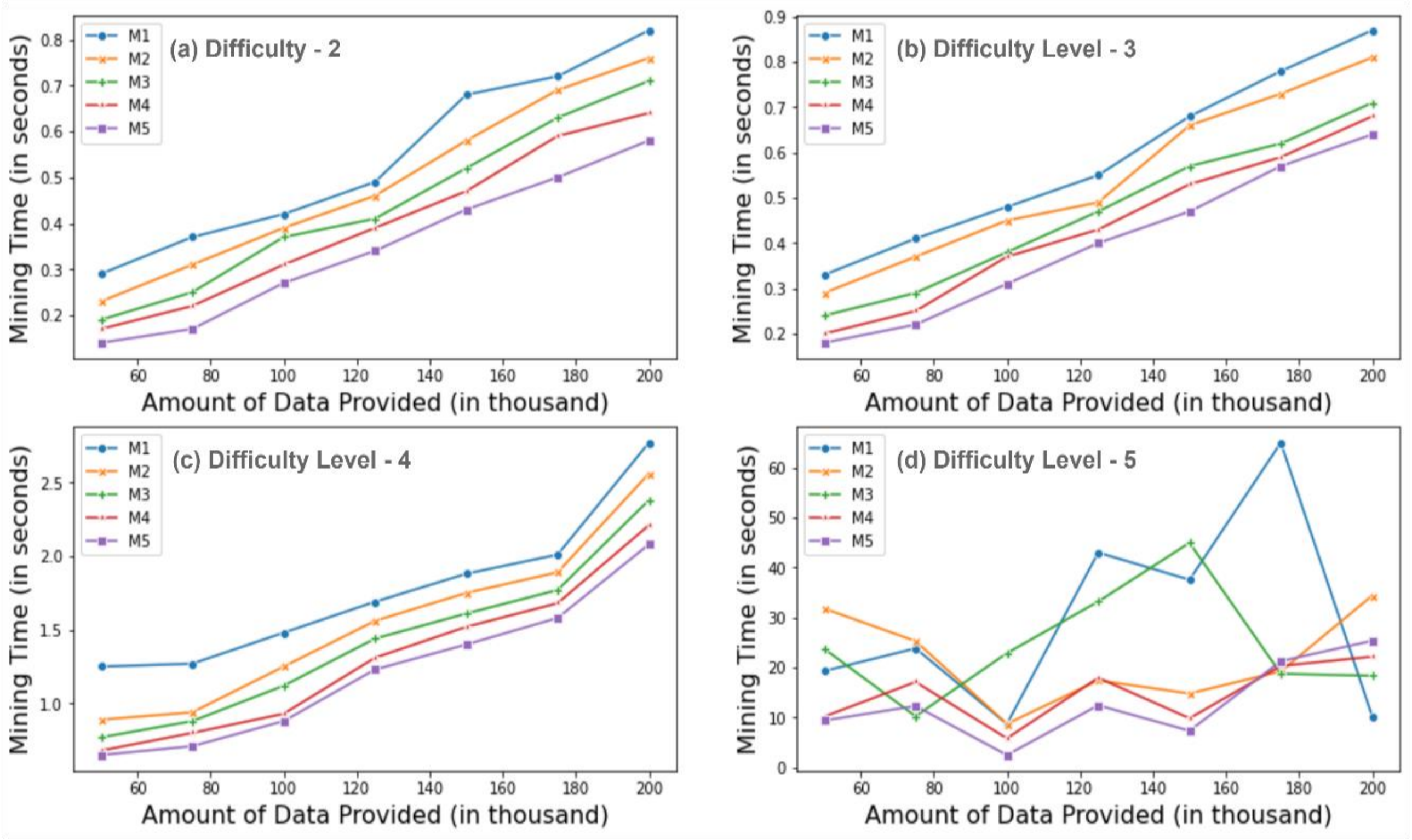

**Figure 14.** Confusion matrix for incremental learning in blockchain network.

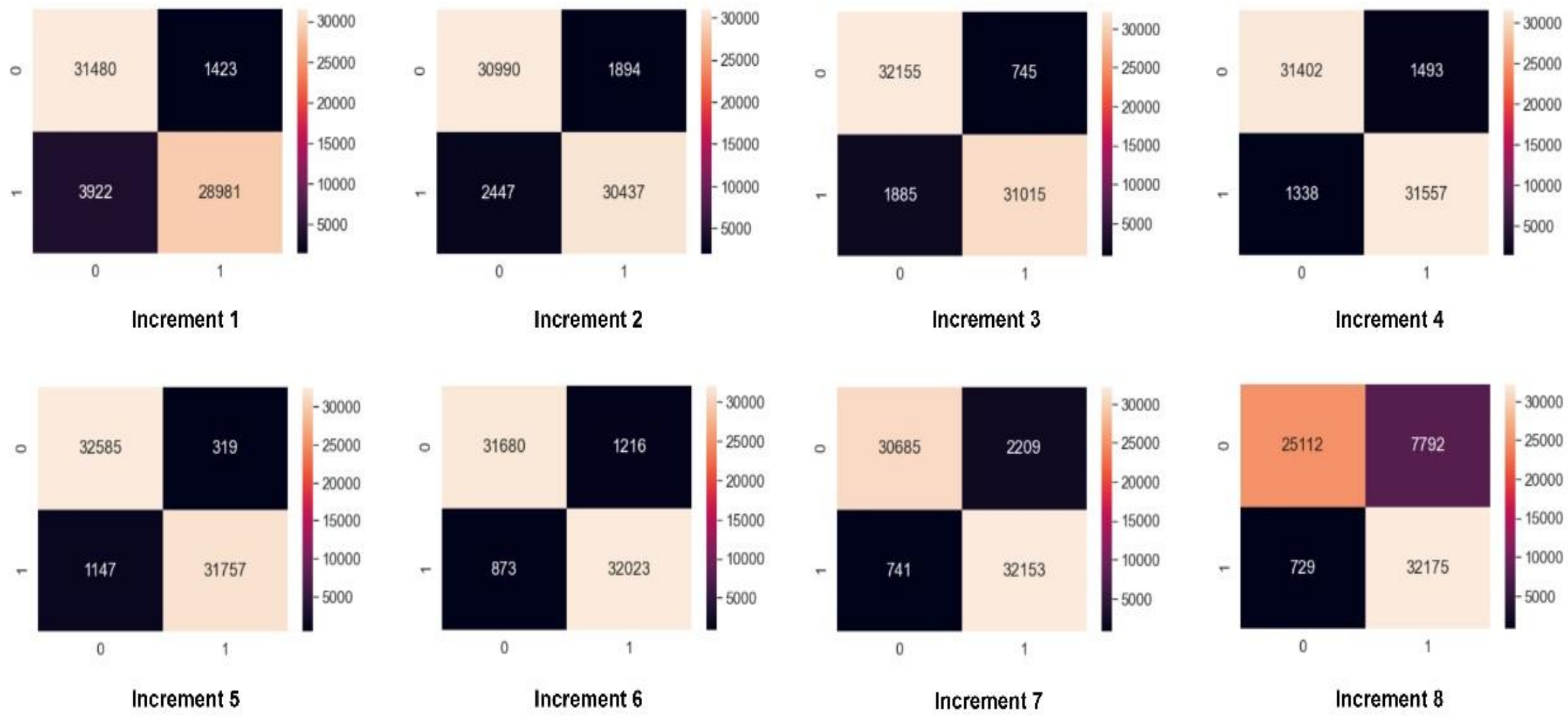

**Figure 15.** **Every parameter update for incremental learning in blockchain network.**

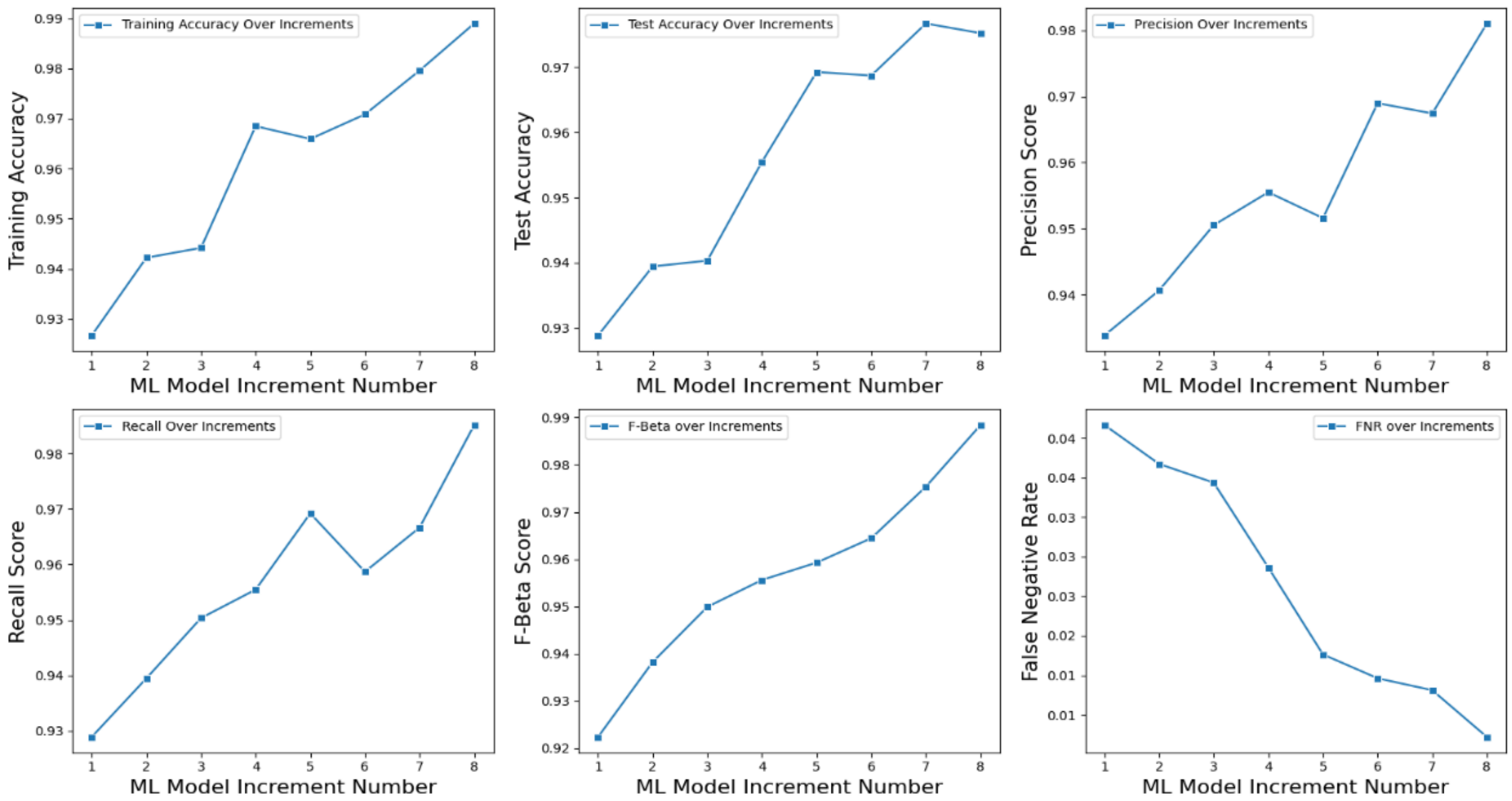

Figure 15 depicts that only increasing Fbeta values were chosen to update the model while keeping the other parameters within the average of their previous results. There is no drastically decreasing or increasing mismatch in training and testing accuracy, which proves that the model is not overfitting over the period of increment. The confusion matrix in figure 14 also advocates for the same speculation. The false-negative rate slowly decreases, while the precision and recall show trade-off behavior. For instance, for increment 5, it is noticeable that a sudden plunge in recall caused a hit to the precision resulting in a sudden decline. The increasing Fbeta threshold forces the recall to increase as well, causing the precision to take hit up to a certain level.

## VII. Discussion

The preceding sections describe and simulate our proposed system. It has been shown that, with collaborative model training, the performance of the ML model can enhance, and the model becomes robust at each iteration in terms of generating fewer false negatives. Where the first model produced 3922 false-negative, with an incremental update to the ML model, the false positive was bought down to 729 for the same amount of data. This section provides detailed theoretical and practical implications of our study.

### *A. Theoretical Implication*

The use of blockchain has enabled our system to secure the privacy of the data by creating a decentralized platform where e-commerce organizations can safely collaborate to incrementally train an ML model without revealing their identity or business data. Data is a crucial factor in highly competitive markets, and e-commerce is undoubtedly a highly competitive market where collaboration appears to be perplexing [67]. On the contrary, transaction fraud causes massive financial loss to these organizations [2]. Machine learning techniques have proven their capability in terms of accuracy and reliability over the last decade with various practical applications. These machine learning models need the actual (real-world) representation of real-life data to train and learn [68]. While on the other hand, transaction data cannot be publicly shared. As a solution to this problem, our system employs a combination of blockchain and smart contract to incrementally and collaboratively train a fraud detection machine learning algorithm without the organizations having to compromise their data and privacy. The data is not shared with the central server. Instead, the data is used within the node itself (off-chain) and only the metrics and results are stored in the central server.

Mining criteria are the collective set of conditions that need to be met before broadcasting a new node [69]. It is an important decision for the efficiency of the overall network [69]. Once new data comes in and partially trains the model, the system decides whether to update the model or not (creating a new block). We have designed a mining criterion specific to our incremental learning problem. Keeping "false-negative rate" as the center of attention, the mining criteria mines a block only if a better false-negative rate is achieved. There might be cases where the false-negative rate improves, but the other metrics are drastically decreasing. To avoid such a scenario, the mining criteria check if the other three metrics (precision, recall and Fbeta) are in the range of the statistical average of the previous best results.

Incentive is greatly important for keeping the participants enthusiast to contribute to the system [70]. An adaptive incentive mechanism makes sure that a fair incentive is given to the rightful participant [71]. A difficulty adaptive incentive mechanism has also been designed to encourage the organizations to participate, which incentivizes the participants based on how difficult it is to update a model. The overall system has been designed in such an order that there are no conflicts between data uploading and data on chaining. Rather than being simultaneous, these two processes happen in sequential order, hence, avoiding the conflict between data on-chaining and uploading. The theoretical implications can be summarized as follows:

- Securing privacy of the organizations by off-chain implementations of machine learning tasks.
- Introducing an adaptive incentive mechanism specifically for an incremental machine learning problem.
- Introducing an optimized mining criteria for incremental learning problems in the context of high-traffic systems like e-commerce.
- Our system has been designed in such a way that computational costs are more distributed in different phases rather than being simultaneous. Our time analysis shows that the system is efficient enough.

### B. Practical Implications

The use of blockchain in our proposed solution keeps the institution's identity protected, untraceable and unretrievable [72]. The learning history of the model is stored in the blockchain, and smart contract automates the process with inherent security provided by the blockchain network. As a result, a safe collaboration between organizations is assured, and the model becomes considerably more precise and rigid in learning from collaborative data. By using our proposed system, organizations can undertake the necessary steps to collaborate for the collective good. Our system proposes a platform that removes the barrier to collaboration between businesses. The practical implications can be summarized as follows:

- Increasing the synergy among organizations to work together for the betterment of technological advancement.
- With collaboratively achieved improvement of current technologies, businesses are expected to grow.
- The solution is generic and can be adopted in any case where collaboration is essential but is bound by commercial bindings.

## VIII. Conclusion

This study proposes an approach where e-commerce organizations can work collaboratively to build robust machine learning algorithms incrementally while safekeeping business strategies and mitigating privacy concerns. To bring robustness to current solutions, authentic data from the real marketplace is required. This study takes advantage of state-of-the-art blockchain technology to build a platform for training fraud detection ML algorithms incrementally and collaboratively while protecting the privacy of contributing organizations. Smart contract has been used in this study to automate the process throughout the system with absolute sturdiness. No functionality within the system is alterable by anyone, and this establishes absolute trust among organizations to share their data safely. An adaptive incentive mechanism has been developed to encourage organizations by providing fair incentives. The reward calculated by the incentive mechanism is positively correlated with the difficulty of improving a model's performance. The more difficult it is to update a model, the more incentive will be disbursed by our system. Authentic data is more likely to impact the model's performance; hence, the real data provider gets highly incentivized.

By employing a throughgoing pre-deploy experiment, we chose passive-aggressive classifier scoring 93.64% testing accuracy as the initial model. From data contributed by the organizations, the model incrementally learns within the blockchain network. The testing accuracy, the Fbeta score, precision and recall reach close to 99% via incremental machine learning. The false-negative rate was bought down to 0.007% from a starting score of 0.04%. The confusion matrix of the increments shows that the false-negative amount decreases to 729 from 3922 as the model learns from new data and improves its capability. We also examined the mining time depending on the blockchain's difficulty level and the different volumes of data provided to the system. The result shows that for difficulty levels 2 and 3, and with a varying amount of data, the mining time is below a second. For difficulty level 3, a maximum of 3 seconds can be taken to mine a block. And for difficulty 5, a much harder problem needs to be solved, and hence, the maximum mining time is approximately 70 seconds.

Our approach is generic to be applied in sectors where data privacy and security are essential, but collaboration among organizations can bring about much better performance and accuracy. For future studies, we intend to build class-adaptive solutions under a stream of data.

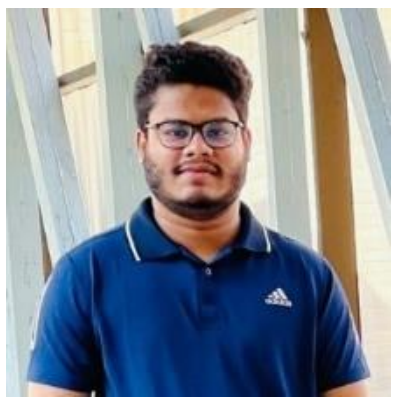

**Tahmid Hasan Pranto** pursued his Bachelor in Computer Science and Engineering from North South University, Dhaka, Bangladesh. Currently, he is working as a Research Assistant at ECE Department of North South University under the supervision of professor Dr. Rashedur M. Rahman. He has a keen interest in emerging technologies like blockchain and machine learning. He has published research works in peer-reviewed journals like PeerJ Computer Science, Applied Artificial Intelligence and Cybernautics and Systems. He has also published in conference proceedings like IWANN, 2021. Current research of him explores the scope and feasibility of artificial intelligence in decentralized platforms by employing incremental machine learning and federated learning approaches in the blockchain-enabled decentralized systems.

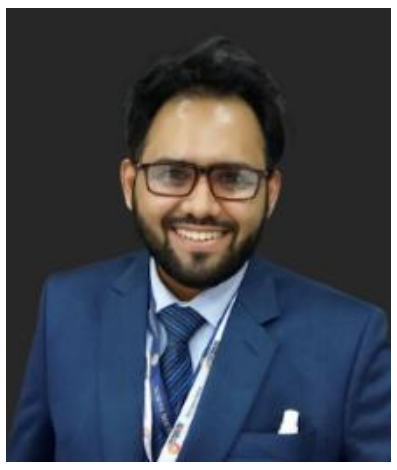

**Kazi Tamzid Akhter Md Hasib** is a Computer Science and Engineering Undergrad from North South University, Dhaka, Bangladesh. He is a Research Assistant at North South University under the supervision of professor Dr. Rashedur M. Rahman. He leads the NSU_Inception team, which participated in various National & International hackathons, including Blockchain Olympiad, NASA Space App Challenges, KIBO Robot Programming Challenges, Microsoft Imagine Cup, AISEC CodeFest, HackNSU (Flagship hackathon of NSU ACM Students Chapters). He achieved Top3 places in some of these hackathons. He is also associated with tech-startup as a founding member. His interests include blockchain, machine learning, deep learning, and computer vision. He has published research work in peer-reviewed journals. His significant interest in Web3.0 has been reflected in his research works. He also has a deep interest in Deep Learning & Computer Vision.

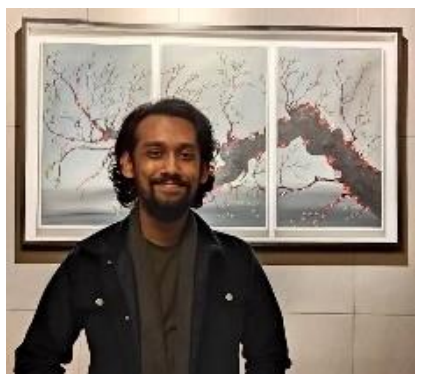

**Tahsinur Rahman** is a Computer Science and Engineering undergraduate student from BRAC University, Dhaka, Bangladesh. He is currently working as a Product Management Intern at HungryNaki, Daraz. Previously he worked as a part-time Data Annotator at Intelligent Machines. His current research interest includes Cybersecurity, Machine Learning, and Blockchain. He has a keen interest in distributed machine learning techniques and decentralized web systems.

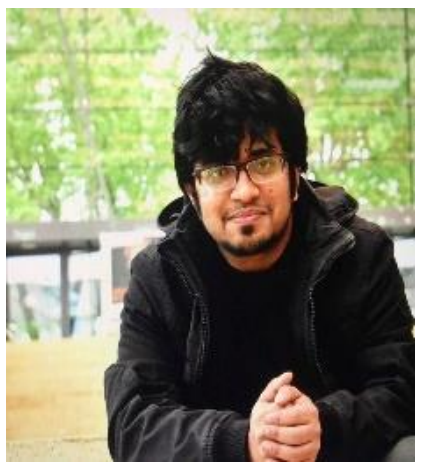

**AKM Bahalul Haque** is a Junior Researcher in the Department of Software Engineering at LUT University. Earlier, he was a lecturer at the Department of Electrical and Computer Engineering, North South University. His works have been accepted and published in international conferences and peer-reviewed journals, including IEEE Access, Expert Systems, Cybernetics and Systems, various international conference proceedings, Tylor and Francis Books, and Springer Book. His research interests include Explainable AI, blockchain, data privacy and protection, and human-computer interaction.

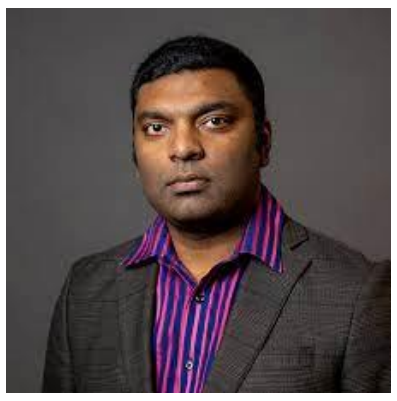

**Dr. A.K.M. Najmul Islam** received his Ph.D. degree in information systems from the University of Turku, Finland, and his M.Sc. (Eng.) degree from the Tampere University of Technology, Finland. He is currently an Adjunct Professor at Tampere University, Finland. He is also an Associate Professor at LUT University, Finland. His research has been published in top outlets, such as IEEE ACCESS, European Journal of Information Systems, Information Systems Journal, Journal of Strategic Information Systems, Technological Forecasting and Social Change, Computers in Human Behavior, Internet Research, Computers & Education, Journal of Medical Internet Research, Information Technology & People, Telematics & Informatics, Journal of Retailing and Consumer Research, Communications of the AIS, Journal of Information Systems Education, AIS Transaction on Human-Computer Interaction, and Behavior & Information Technology. His research interest includes human–cantered computing.

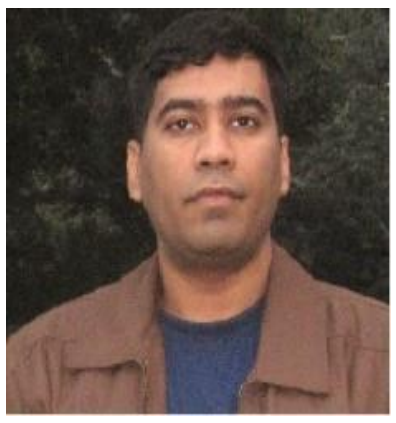

**Dr. RASHEDUR M. RAHMAN** works as a Professor in the Electrical and Computer Engineering Department at North South University, Dhaka, Bangladesh. He received his Ph.D. from the University of Calgary, Canada, in 2007 and an MS degree from the University of Manitoba, Winnipeg, Canada, in 2003. He published more than 150 research papers in the area of parallel and distributed computing, cloud and grid computing, data and knowledge engineering. He is also on the editorial committee of many international journals. His current research interest is in cloud load characterization, VM consolidation, and the application of data mining and fuzzy logic in different decision-making problems.